# Analysis of Implementation of HIEROCRYPT–3 algorithm (and its comparison to CAMELLIA algorithm) using ALTERA devices.

Marcin ROGAWSKI
Military University of Technology
Institute of Mathematics and Cryptology
Faculty of Cybernetics
mailto:mrogawski@poczta.onet.pl

**June 2003**

Table of contents:

**Introduction**

Alghoritms: HIEROCRYPT-3, CAMELLIA and ANUBIS, GRAND CRU, NOEKEON, NUSH, Q, RC6, SAFER$_{++128}$, SC2000, SHACAL were requested for the submission of block ciphers (*high level block cipher*) to NESSIE (*New European Schemes for Signatures, Integrity, and Encryption*) project. The main purpose of this project was to put forward a portfolio of strong cryptographic primitives of various types. The NESSIE project was a three year long project and has been divided into two phases. The first was finished in June 2001r. CAMELLIA, RC6, SAFER$_{++128}$ and SHACAL were accepted for the second phase of the evaluation process.

HIEROCRYPT-3 had key schedule problems [5, 7], and there were attacks for up to 3,5 rounds out of 6 [1, 3, 7], at least hardware implementations of this cipher were extremely slow [12]. HIEROCRYPT-3 was not selected to Phase II.

CAMELLIA was selected as an algorithm suggested for future standard [10].

In the paper we present the hardware implementations these two algorithms with 128-bit blocks and 128-bit keys, using ALTERA devices and their comparisons.

## 1. Short description of the HIEROCRYPT-3 cipher

The HIEROCRYPT-3 block cipher algorithm was designed by TOSHIBA Corporation and its detailed specification is given in [11]. We have implemented the version of the algorithm with 128 bit blocks and 128 bit main key. The HIEROCRYPT-3 has 6 rounds and each round needs two 128 bit subkeys and one 128 bit subkey is necessary to EXOR with the text block at the end of the encryption process.

Structure of HIERCORYPT-3 cipher is based on "wide trail strategy" described by Joan Deamen in his PhD in 1995 [4]. This paper suggested design strategies based on linear and differential cryptoanalysis. In HIEROCRYPT-3: non-linearlity is represented by two layers (2x16 simultaneously working sboxes) and linear layers are represented by matrices: MDS$_L$(operating on 4x32-bit word) and MDS$_H$ (operating on 128-bit block of data). In this strategy, obviously, each round of encryption and decryption process is dependent on subkeys (In HIEROCRYPT-3: twice EXOR with 2x128-bit subkey).

## 1.1 Round of encryption

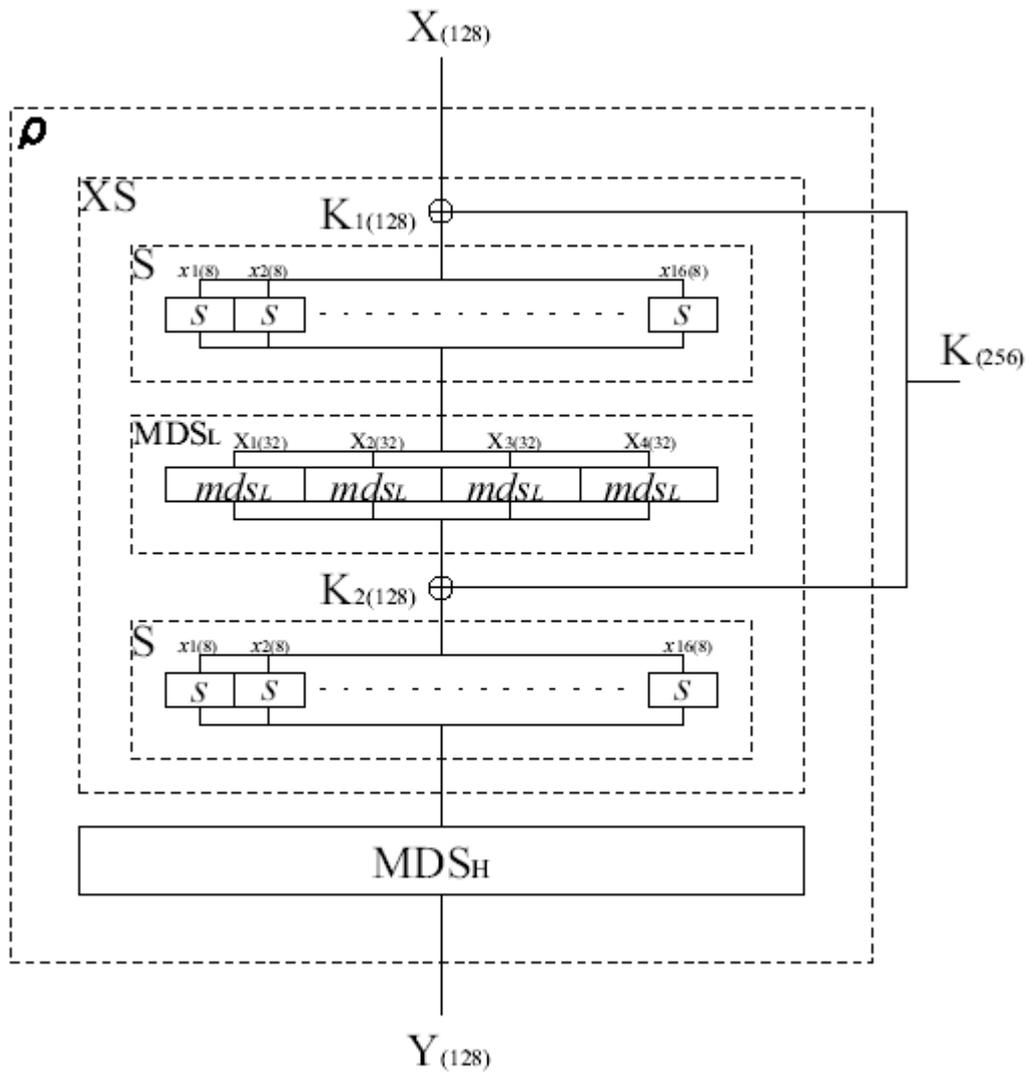

**Fig. 1.1: Round of encryption**

$K_{1(128)}$ – first 128-bit round key

$K_{2(128)}$ – second 128-bit round key

S – substitution box

$mds_L$ – MDS lower level

$MDS_H$ – MDS higher level

XS – last round of encryption process (without $MDS_H$).

## 1.2 Key Schedule

The main part of key scheduling consists of the intermediate key generation part and the round key generation part, preceded by the intermediate key initialization. The intermediate key part recursively generates intermediate key outputs $Z^{(t)}_{(256)}$ (t = 1, 2,..., T+1), and the round key generation part generates round keys $K^{(t)}_{(256)}$ (t = 1, 2,..., T+1) from the corresponding intermediate keys.

Notations:

K – main key,

Z – intermediate key,

$G^{(t)}_{(64)}$ – constant for n-round,

$X_{(n)}$ – input binary value of size n,

$Y_{(n)}$ – output binary value of size n,

$K_{1(64)} \| K_{2(64)} \| K_{3(64)} \| K_{4(64)} = K_{(128)}$ (Main key consists of four 64-bits values),

$Z_{1(64)} \| Z_{2(64)} \| Z_{3(64)} \| Z_{4(64)} = Z_{(128)}$ (Intermediate key consists of four 64-bits values),

$t_{turn}$ – 4 for 128/192 bit main key, 5 for 256 bit key,

t – round number.

### 1.2.1 Intermediate key generation procedure

Iterative update of the intermediate key σ (1<= t <= $t_{turn}$)

$$Z^{(t)}_{(256)} = \sigma(Z^{(t-1)}_{(256)}, G^{(t)}_{(64)}),$$

definition of function:

$$W^{(t-1)}_{1(64)} \| W^{(t-1)}_{2(64)} = P^{(32)}(Z^{(t-1)}_{1(64)} \| Z^{(t-1)}_{2(64)})$$

$$Z^{(t)}_{1(64)} = Z^{(t-1)}_{2(64)},$$

$$Z^{(t)}_{2(64)} = Z^{(t-1)}_{1(64)} \oplus F_\sigma(Z^{(t-1)}_{2(64)} \oplus Z^{(t)}_{3(64)}),$$

$$Z^{(t)}_{3(64)} = M_{5E}(W^{(t-1)}_{1(64)}) \oplus G^{(t)}_{(64)},$$

$$Z^{(t)}_{4(64)} = M_{5E}(W^{(t-1)}_{2(64)}).$$

Iterative update of the intermediate key $\sigma^{-1}$ ($t_{turn} +1 <= t <= T+1$)

$$Z^{(t)}_{(256)} = \sigma^{-1}(Z^{(t-1)}_{(256)}, G^{(t)}_{(64)}),$$

definition of function:

$$Z^{(t)}_{1(64)} = Z^{(t-1)}_{2(64)} \oplus F_\sigma(Z^{(t-1)}_{1(64)} \oplus Z^{(t-1)}_{3(64)}),$$
$$Z^{(t)}_{2(64)} = Z^{(t-1)}_{1(64)},$$
$$W^{(t)}_{1(64)} = M_{B3}(Z^{(t-1)}_{3(64)} \oplus G^{(t)}_{(64)}),$$
$$W^{(t)}_{2(64)} = M_{B3}(Z^{(t-1)}_{4(64)}),$$
$$Z^{(t)}_{3(64)} \| Z^{(t)}_{4(64)} = P^{(32)-1}(W^{(t-1)}_{1(64)} \| W^{(t-1)}_{2(64)}).$$

### 1.2.2 Round key generation procedure

Key generation procedure for $1 \leq t \leq t_{turn}$.

$$V^{(t)}_{(64)} = F_\sigma(Z^{(t-1)}_{2(64)} \oplus Z^{(t-1)}_{3(64)}),$$
$$K^{(t)}_{1(64)} = Z^{(t-1)}_{1(64)} \oplus V^{(t)}_{(64)},$$
$$K^{(t)}_{2(64)} = Z^{(t)}_{3(64)} \oplus V^{(t)}_{(64)},$$
$$K^{(t)}_{3(64)} = Z^{(t)}_{4(64)} \oplus V^{(t)}_{(64)},$$
$$K^{(t)}_{4(64)} = Z^{(t-1)}_{2(64)} \oplus Z^{(t)}_{4(64)},$$

Key generation procedure for $t_{turn}+1 \leq t \leq T+1$.

$$V^{(t)}_{(64)} = F_\sigma(Z^{(t-1)}_{1(64)} \oplus Z^{(t)}_{3(64)}),$$
$$K^{(t)}_{1(64)} = Z^{(t)}_{1(64)} \oplus Z^{(t-1)}_{3(64)},$$
$$K^{(t)}_{2(64)} = W^{(t)}_{1(64)} \oplus V^{(t)}_{(64)},$$
$$K^{(t)}_{3(64)} = W^{(t)}_{2(64)} \oplus V^{(t)}_{(64)},$$
$$K^{(t)}_{4(64)} = Z^{(t-1)}_{1(64)} \oplus W^{(t)}_{2(64)},$$

### 1.2.3 Table of key schedule (128 – bit main key)

|  | t | operacja | $G^{(t)}_{(64)}$ |
|---|---|---|---|
| - | -1 (PAD) | - | $H_3 \| H_2$ |
| - | 0 (PW) | $\sigma_0$ | $G_0(5)$ |
| $K^{(1)}_{(256)}$ | 1 | $\sigma$ | $G_0(0)$ |
| $K^{(2)}_{(256)}$ | 2 | $\sigma$ | $G_0(1)$ |
| $K^{(3)}_{(256)}$ | 3 | $\sigma$ | $G_0(2)$ |
| $K^{(4)}_{(256)}$ | 4 | $\sigma$ | $G_0(3)$ |
| $K^{(5)}_{(256)}$ | 5 | $\sigma^{-1}$ | $G_0(3)$ |
| $K^{(6)}_{(256)}$ | 6 | $\sigma^{-1}$ | $G_0(2)$ |
| $K^{(7)}_{(256)}$ | 7 | $\sigma^{-1}$ | $G_0(1)$ |

**Table 1.1: Key Schedule**

PAD – padding, this operation extends various length main keys to the 256-bit size.

PW – key pre-whitening.

$K^{(1)}_{(256)}$ - $K^{(6)}_{(256)}$ – subkeys for rounds of encryption

$K^{(7)}_{(256)}$ – subkey for AK operation.

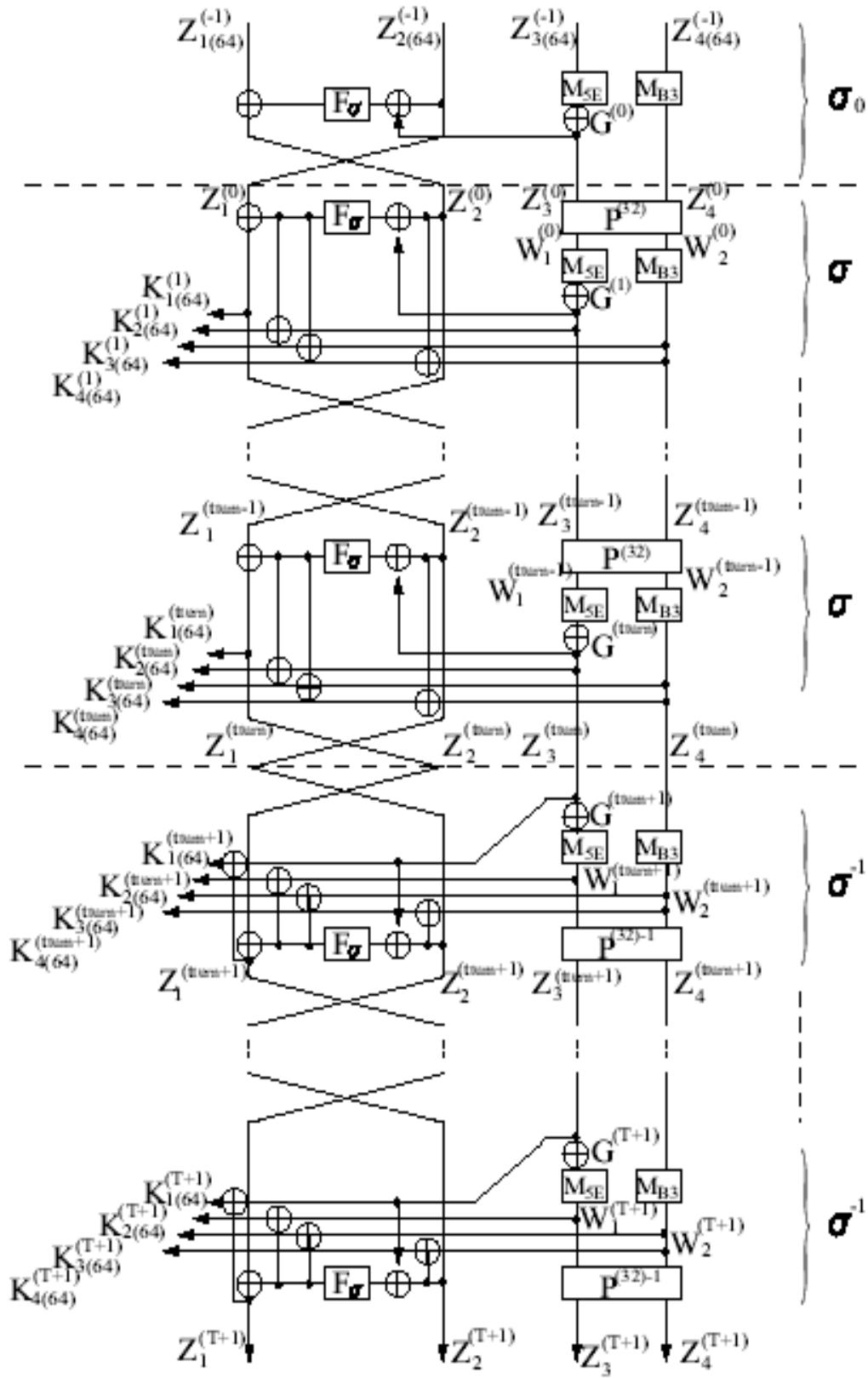

**Fig. 1.2: Key schedule**

## 1.3  Encryption

The T-round encryption of Hierocrypt–3 consists of (T-1) operations of round function ρ, an operation of XS-function, and the final key addition (AK).

$$P_{(128)} \equiv X^{(0)}_{(128)} \xmapsto{\rho} X^{(1)}_{(128)} \xmapsto{\rho} \cdots \xmapsto{\rho} X^{(T-1)}_{(128)} \xmapsto{XS} X^{(T)}_{(128)} \xmapsto{AK} C_{(128)}$$

T is 6,7, or 8 for 128-, 192-, or 256-bit, respectively.

The 128-bit value $X^{(i)}_{(128)}$ is the output of the i-th operation of round function ρ (i = 1, 2,..., T-1). The plaintext $P_{(128)}$ is assigned to the 0-th value $X^{(0)}_{(128)}$. The value $X^{(t)}_{(128)}$ is the output of the t-th operation of ρ-function for the input $X^{(t-1)}_{(128)}$ and the round key $K^{(t)}_{(256)}$.

$$X^{(t)}_{(128)} = \rho(X^{(t-1)}_{(128)}, K^{(t)}_{(256)}), \quad t = 1, 2, \cdots, T-1.$$

Similarly, $X^{(T)}_{(128)}$ is the output of XS-function for the input $X^{(T-1)}_{(128)}$ and the final key $K^{(T)}_{(256)}$.

$$X^{(T)}_{(128)} = XS(X^{(T-1)}_{(128)}, K^{(T)}_{(256)}).$$

The ciphertext $C_{(128)}$ is given as the addition (XOR, exclusive or) between the T-th round output $X^{(T)}_{(128)}$ and the first half of the final key $K^{(T+1)}_{1(128)}$.

$$C_{(128)} = X^{(T)}_{(128)} \oplus (K^{(T+1)}_{1(64)} \| K^{(T+1)}_{2(64)}).$$

## 1.4  Decryption

The decryption of Hierocrypt–3 is the inverse of encryption, and consists of the final key addition, the inverse of XS-function (XS$^{-1}$), and (T-1) inverse operations of round function (ρ$^{-1}$).

$$\begin{aligned} X^{(T)}_{(128)} &= C_{(128)} \oplus (K^{(T+1)}_{1(64)} \| K^{(T+1)}_{2(64)}), \\ X^{(T-1)}_{(128)} &= XS^{-1}(X^{(T)}_{(128)}, K^{(T)}_{(256)}), \\ X^{(t-1)}_{(128)} &= \rho^{-1}(X^{(t)}_{(128)}, K^{(t)}_{(256)}), \quad t = T-1, \ldots, 2, 1. \end{aligned}$$

The plaintext $P_{(128)}$ is given as the final output $X^{(0)}_{(128)}$.

$$P_{(128)} = X^{(0)}_{(128)}.$$

## 2. Analysis of the HIEROCRYPT-3 main components

The analysis presented in this section concerns the ability of implementing HIEROCRYPT-3 using ALTERA FPGA devices. In the following section we will discuss: all basic functions used in the algorithm, and the way of implementing them in ALTERA FPGA. These basic functions include:
- s-boxes,
- MDS lower level,
- MDS higher level,
- $P^{(n)}$ – function,
- $M_{5E}$ – function,
- $M_{B3}$ – function,
- Fσ – function.

### 2.1 Round of encryption and decryption

#### 2.1.1 Substitution boxes

Basically, there are two possible ways of implementing s-boxes:
- as a direct logic implementation, or
- as a 2048-bit configured embedded array block (EAB).

We analyzed both solutions (there are 40 sboxes in the HIEROCRYPT-3: 32 in round of encryption or decryption and 8 in key schedule and FLEX10KE have only 24 EABs). The best solution seems to be the implementation:
- one layer of sboxes from round of encryption (16 sboxes) and 8 sboxes from key schedule implemented in EABs (24 sboxes together),
- one layer of sboxes from round of encryption (16 sboxes) implemented as a direct logic implementation. We used DAMAIN tool [13], developed at Warsaw University of Technology, for the functional decomposition of sbox (it provides more efficient and faster implementation than Max PLUS optimalisation methods).

### 2.1.2 MDS lower level

Implementation of the MDS matrix can seem very difficult, but closer analysis of operation performed in this matrix leads us to different conclusion.

$$\begin{bmatrix} Y_{1(8)} \\ Y_{2(8)} \\ Y_{3(8)} \\ Y_{4(8)} \end{bmatrix} = \begin{bmatrix} C4 & 65 & C8 & 8B \\ 8B & C4 & 65 & C8 \\ C8 & 8B & C4 & 65 \\ 65 & C8 & 8B & C4 \end{bmatrix} * \begin{bmatrix} x_{1(8)} \\ x_{2(8)} \\ x_{3(8)} \\ x_{4(8)} \end{bmatrix}$$

**Table 2.1: MDS lower level**

$$y_{1(8)} = C4*x_{1(8)} \oplus 65*x_{2(8)} \oplus C8*x_{3(8)} \oplus 8B*x_{4(8)}$$
$$y_2(8) = 8B*x_1(8) \oplus C4*x_2(8) \oplus 65*x_3(8) \oplus C8*x_4(8)$$
$$y_3(8) = C8*x_1(8) \oplus 8B*x_2(8) \oplus C4*x_3(8) \oplus 65*x_4(8)$$
$$y_4(8) = 65*x_1(8) \oplus C8*x_2(8) \oplus 8B*x_3(8) \oplus C4*x_4(8)$$

primitive polynomial for this field $x^8 + x^6 + x^5 + x + 1$.

Each 32-bit input value consists of four 8-bit values. Each 8-bit value is multiplied by a vector from the matrix and the results of all multiplication in each row of the MDS matrix are finally XORed bit by bit.

Implementation of multiplication by C4h:

OUT[7..0] = C4 * IN[7..0]
IN[7..0] – input 8-bit value from $GF(2^8)$,
(IN[7] is the most significant bit in input value, IN[0] is the less significant bit in input value)
OUT[7..0] – output 8-bit value from $GF(2^8)$
(OUT[7] is the most significant bit in output value, OUT[0] is the less significant bit in output value)

$$OUT[7] = IN[7] \oplus IN[2] \oplus IN[1] \oplus IN[0];$$
$$OUT[6] = IN[6] \oplus IN[1] \oplus IN[0];$$
$$OUT[5] = IN[5] \oplus IN[2] \oplus IN[1];$$
$$OUT[4] = IN[7] \oplus IN[4] \oplus IN[2];$$
$$OUT[3] = IN[6] \oplus IN[3] \oplus IN[1];$$
$$OUT[2] = IN[5] \oplus IN[2] \oplus IN[0];$$
$$OUT[1] = IN[4] \oplus IN[1];$$
$$OUT[0] = IN[3] \oplus IN[2] \oplus IN[1];$$

We implement the rest of multiplication (by C8h, 65h, 8Bh) in the same way.

### 2.1.3 MDS higher level

This kind of operation is easily implemented in hardware. This linear transformation operates on 128-bit block. It consists of sixteen input 8-bit values. The only thing we have to do is to XOR correct 8-bit values. In this way we receive all of sixteen output 8-bit values.

## 2.2 Key schedule

### 2.2.1 $P^{(n)}$ – function

This kind of operation is easily implemented in hardware, too. This linear transformation operates on n-bit block. It consists of four input n-bit values. The only thing we have to do is to XOR correct n-bit values. In this way we receive four output n-bit values. We executed this transformation similarly to MDS higher level operation.

### 2.2.2 $M_{5E}$ – function

This kind of operation is easily implemented in hardware. This linear transformation operates on 64-bit block. It consists of eight input 8-bit values. The only thing we have to do is to XOR correct 8-bit values. In this way we receive eight output 8-bit values. We implemented this transformation similarly to MDS higher level operation.

### 2.2.3 $M_{B3}$ – function

This transformation is very similar to $M_{5E}$. We implemented this in the same way.

## 2.2.4 Fσ – function

We described the ways of implementation of each part of this transformation in previous units. Sboxes in 2.1.1 and $P^{(n)}$ – function in 2.2.1.

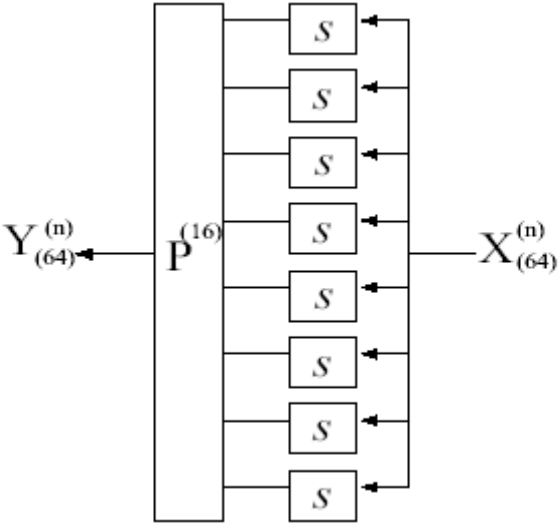

**Fig 2.1: Fσ – function**

# 3. Implementation of the HIEROCRYPT-3 algorithm

## 3.1 Designing criteria – main assumptions of implementation

TOSHIBA Corp. prepared hardware implementation in ALTERA Max+Plus II using Flex 10K devices. This FPGA family is considered to be correct for dedicated cryptographic solutions. In this paper we will also try to present the realisation of HIEROCRYPT-3 using these ALTERA products (Max+plus II and Flex 10 KE).

There are various approaches to the realisation block ciphers. Some papers suggest that the generation of the subkeys and the round calculations should be parally executed. In the first order the subkey to the round number one is calculated. Then the round is executed and the subkey to the next round is calculated. The main adventage of this design is the fact that we do not need to store the subkeys, they are currently calculated [6].

Another proposition is to implement the key generation algorithm in other units of the cipher. In the first phase (called key setup) all necessary subkeys are generated and they are stored in the internal implemented registers (or memory). There is in the second phase encryption or decryption process only [2].

Both realisations have got lots of advantages and disadvantages and both are most suitable for algorithms presented in these papers [2, 6]. Many other implementations of algorithms [14] presented in these papers [2, 6] proved correctness of the choices made.

Conclusion: The best realisation of LOOP (iterative) architecture depends on special features of every symmetric algorithm.

Our proposals of HIEROCRYPT-3 implementation use the features of both realisations. There are two phases of working FPGA circuit: key setup phase and encryption (decryption) phase. In the first order we realise the precomputation of some parts of subkeys algorithm generation (details are described in the units concerning the implementation of each project). This phase is finished when "high state" (logical one) appears on the output READY. Logical one remains on the output READY till there is a rising edge on the input RESET. If logical zero is on the output READY, every rising edge on the input START is ignored. If logical one is on the output READY, the first rising edge on the input START begins the work of the unit. During the work of the unit logical one is on the output WORK. After the end of the encryption (decryption) process on the output WORK appears logical zero.

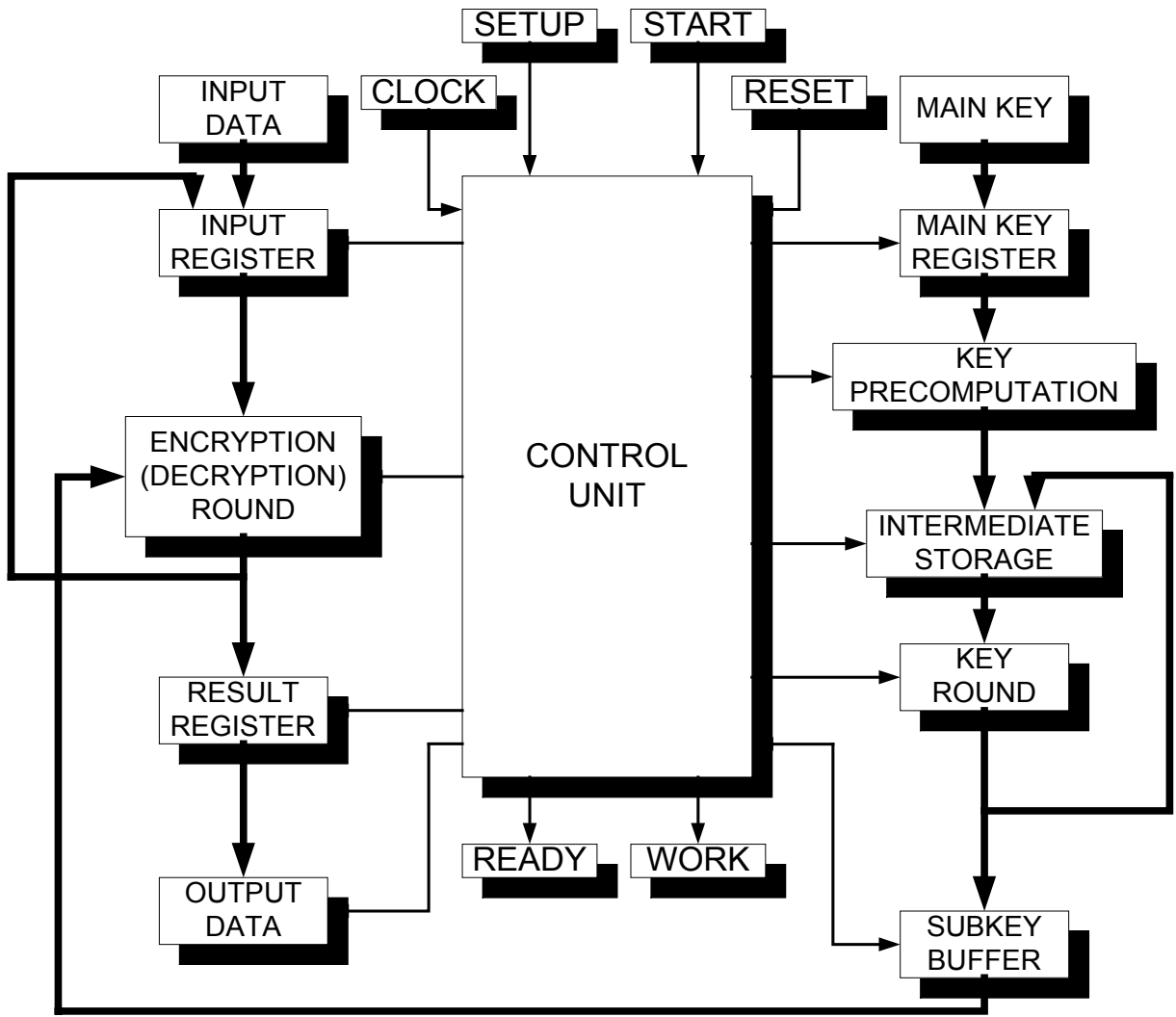

**Fig. 3.1: The encryption (decryption) unit.**

## 3.2 TOSHIBA Corp. project and its results

TOSHIBA Corp. presented the performance of their FPGA projects in the paper "Self evaluation: HIEROCRYPT-3" [12]. They prepared two different solutions: High Speed Implementation and Small Area Implementation.

|  | Logic elements | Throughput | Datails |
|---|---|---|---|
| High speed | 22700 | 52,6 Mb/s | 5 devices !!! |
| Small area | 6300 | 4,1 Mb/s |  |

**Table 3.1: Results of implementation of HIEROCRYPT-3 by TOSHIBA Corp.**

## 3.3 Implementation of HIEROCRYPT-3 with short setup

In the first order we present a solution with short setup. In the phase of key setup, we realise the operation of main key pre-whitening and calculation of the subkey to the round number one. Next, in the external register from INTERMEDIATE STORAGE (Fig. 3.1) there is the first intermediate subkey. The subkey for round number one is in SUBKEY BUFFER. Encryption (decryption) unit is ready (READY – logical one) for working. During the work (WORK – logical one) of encryption (decryption) unit the generation of the subkeys and the round calculations are parallely executed. After the end of encryption (decryption) process, there is the first intermediate subkey in the external register from INTERMEDIATE STORAGE and subkey for round number one. In the last clock cycle of encryption (decryption) process there is an operation which writes these values from other registers. The unit is ready (READY is logical one, until the rising edge on the input RESET shows) to encrypt (decrypt) next 128-bit block.

The first feature of all implemented projects (3.3, 3.4, 3.5) is the way of sbox realisation. It is described in 2.1.1 and it is caused by technology of Flex 10 KE (There are 24 blocks of EAB). For the first time of the implementation process the whole project fits to the one FPGA circuit. Its resources (logic elements and memory bits) are enough for all project elements.

This project executes correct encryption (decryption) process during 8 clock cycles. Frequency of the clock could be 8,05 MHz and the throughput of this project is 115 Mb/s.

The key round is the critical path.

## 3.4 Implementation of HIEROCRYPT-3 with long setup

The next feature of algorithm HIEROCRYPT-3 considered in the project with long setup is symmetry in the intermediate part of the key schedule (table 1.1). Using iterative update of intermediate key $\sigma$ described in section 1.2.1. we can compute 1600 bits (5x256-bit intermediate subkeys and five 64-bit outputs of $F_\sigma$ - function for corresponding intermediate subkeys), which are the critical data to the computation of all round subkeys. These 1600 bits are stored in external registers in INTERMEDIATE STORAGE. In this case we do not need to implement some parts of the algorithm (for example: $\sigma^{-1}$ operation). These pre-computations are executed in the setup. Logical one appears on the output READY, FPGA circuit is ready to encryption (decryption) process.

During the work (WORK – logical one) of encryption (decryption) unit the generation of the subkeys and the round calculations are parally executed. Subkeys for the next round are calculated using the 1600 bits from INTERMEDIATE STORAGE. These stored bits are on the input to the key round (described in 1.2.2).

This project executes correct encryption (decryption) process during 8 clock cycles. Frequency of the clock could be 11,91 MHz and throughput of this project is 190 Mb/s.

The computation of $F_\sigma$ - function output data is critical path.

## 3.5 Implementation of HIEROCRYPT-3 with very long setup

Two very important changes in project with long setup were made and these are the main features of this project.

Firstly, two operations from separated clock cycles became one: operation of XS-function from 7th clock cycle and AK-operation from 8th clock cycle (encryption process takes only 7 clock cycles, instead of 8 from the previous project) are executed in 7th clock cycles.

Second change caused the setup to become longer and encryption (decryption) become shorter. Each iterative update of intermediate key $\sigma$ is executed in three clock cycles.

First clock cycle:
$$W^{(t-1)}_{1(64)} \| W^{(t-1)}_{2(64)} = P^{(32)}(Z^{(t-1)}_{1(64)} \| Z^{(t-1)}_{2(64)})$$

Second clock cycle:
$$Z^{(t)}_{3(64)} = M_{5E}(W^{(t-1)}_{1(64)}) \oplus G^{(t)}_{(64)},$$
$$Z^{(t)}_{4(64)} = M_{5E}(W^{(t-1)}_{2(64)}).$$

Third clock cycle:
$$Z^{(t)}_{1(64)} = Z^{(t-1)}_{2(64)},$$
$$Z^{(t)}_{2(64)} = Z^{(t-1)}_{1(64)} \oplus F_\sigma(Z^{(t-1)}_{2(64)} \oplus Z^{(t)}_{3(64)}),$$

All these operations have almost the same delay – about 28 ns.
This division of this procedure provide that the round of encryption (decryption) is the critical path.

This project executes correct encryption (decryption) process during 7 clock cycles. Frequency of the clock could be 15,64 MHz and throughput of this project is 304 Mb/s.

## 3.6 Extensive implementation of HIEROCRYPT-3

We implement this implementation in STRATIX circuit available in QUARTUS II, because it was not possible to implement it in Flex 10 KE (too much resources necessary). Only one change was executed in the last project. This change was mathematical and it resulted from flexibility of HIEROCRYPT-3 algorithm.

Sbox in HIEROCRYPT-3 is 8x8 size and it is the bijective function, that means it is permutation of $GF(2^8)$ elements. Multiplication by MDS lower level matrix is executed in this way:

$$y_{1(8)} = C4*x_{1(8)} \oplus 65*x_{2(8)} \oplus C8*x_{3(8)} \oplus 8B*x_{4(8)}$$

$$y2(8) = 8B*x_1(8) \oplus C4*x_2(8) \oplus 65*x_3(8) \oplus C8*x_4(8)$$

$$y3(8) = C8*x_1(8) \oplus 8B*x_2(8) \oplus C4*x_3(8) \oplus 65*x_4(8)$$

$$y4(8) = 65*x_1(8) \oplus C8*x_2(8) \oplus 8B*x_3(8) \oplus C4*x_4(8)$$

(primitive polynomial for this field $x^8 + x^6 + x^5 + x + 1$).

Each element from $GF(2^8)$ is firstly multiplied by four constants: C4h, 8Bh, C8h, 65h, and then they are EXORed. Multiplication of all elements from $GF(2^8)$ by constant causes permutation of the elements from $GF(2^8)$.

Hence, we can consider sbox as a permutation. We can consider the multiplication by MDS lower level matrix as a four permutation of each 8 bit value (there are 16 these values in 128 bit block). It is possible to connect sbox with each multiplication and to receive one bijective sbox. MDS lower level matrix is circulant and this feature gives us four classes of new sboxes. Each class is represented by sixteen sboxes (implication from 16 eight bit values). This connection cause that the operation of round encryption (decryption) take only 46ns.

This project execute correct encryption (decryption) process during 7 clock cycles. Frequency of the clock could be 21,73 MHz and throughput of this project is 397 Mb/s.

## 3.7 Summary of HIEROCRYPT-3 cipher implementation.

|  | Efficiency | throughput |
|---|---|---|
| Our proposal: Project with short setup | 8599 LE / 48kb EAB | 115 Mb/s |
| Our proposal: Project with long setup | 9497 LE / 48kb EAB | 190 Mb/s |
| Our proposal: Project with very long setup | 9758 LE / 48kb EAB | 304 Mb/s |
| Our proposal: Extensive solution | 25811 LE | 397 Mb/s |
| TOSHIBA Corp.:high speed | 22700 LE | 52,6 Mb/s |
| TOSHIBA Corp.:small area | 6300 LE | 4,1 Mb/s |

**Table 3.2: Summary of implementation of cipher HIEROCRYPT-3.**

# 4. CAMELLIA algorithm and its implementation

The CAMELLIA block cipher algorithm was designed by NTT Corporation and Mitsubishi Electric Corporation and its detailed specification is given in [8]. We have implemented the version of the algorithm with 128 bit blocks and 128 bit main key. The CAMELLIA has 18 rounds and each round needs one 64 bit subkey. Four 64 bit subkeys are necessary to pre-whitening and post-whitening operations at the beginning and end of the encryption process.

## 4.1 Structure of CAMELLIA algorithm

### 4.1.1 Encryption and decryption round

The data randomizing part has an 18-round Feistel structure with two $FL/FL^{-1}$-function layers after the 6-th and 12-th rounds and 128-bit XOR operations before the first round and after the last round. The key schedule part generates subkeys $kw_{t(64)}$ (t = 1, 2, 3, 4), $k_{u(64)}$ (u = 1, 2, . . . , 18) and $kl_{v(64)}$ (v = 1, 2, 3, 4) from the secret key K. Section 4.1.2 describes in detail the key schedule part.

In the data randomizing part, first the plaintext $M_{(128)}$ is XORed with $kw_{1(64)}||kw_{2(64)}$ and separated into $L_{0(64)}$ and $R_{0(64)}$ of equal length,

$$M_{(128)} \oplus (kw_{1(64)}||kw_{2(64)}) = L_{0(64)}||R_{0(64)}.$$

Then, the following operations are perfomed from r = 1 to 18, except for r = 6 and 12;

$$L_r = R_{r-1} \oplus F(L_{r-1}, k_r),$$
$$R_r = L_{r-1}.$$

For r = 6 and 12, the following is carried out;

$$L'_r = R_{r-1} \oplus F(L_{r-1}, k_r),$$
$$R'_r = L_{r-1},$$
$$L_r = FL(L'_r, kl_{2r/6-1}),$$
$$R_r = FL^{-1}(R'_r, kl_{2r/6}).$$

Lastly, $R_{18(64)}$ and $L_{18(64)}$ are concatenated and XORed with $kw_{3(64)}||kw_{4(64)}$. The resultant value is the ciphertext, i.e.,

$$C_{(128)} = (R_{18(64)}||L_{18(64)}) \oplus (kw_{3(64)}||kw_{4(64)}).$$

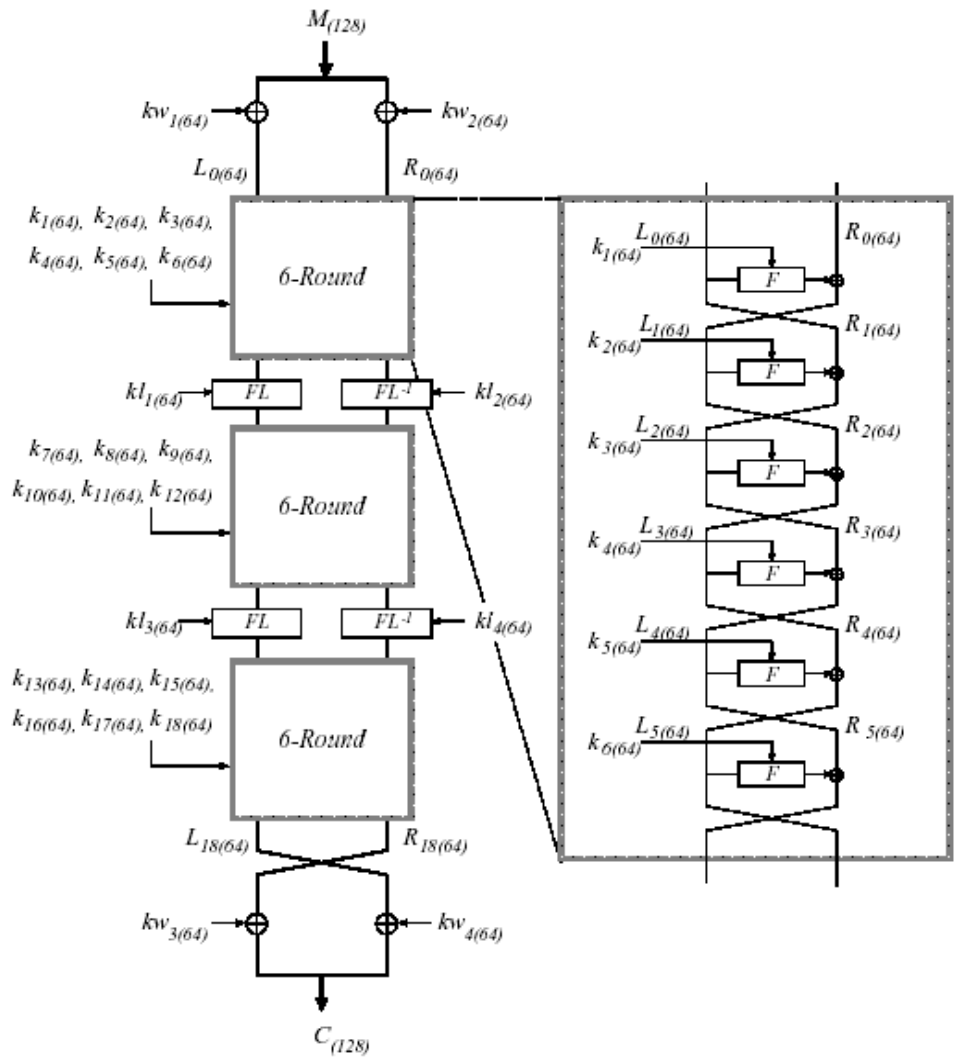

**Fig. 4.1: Encryption procedure of CAMELLIA version 128-bit main key.**

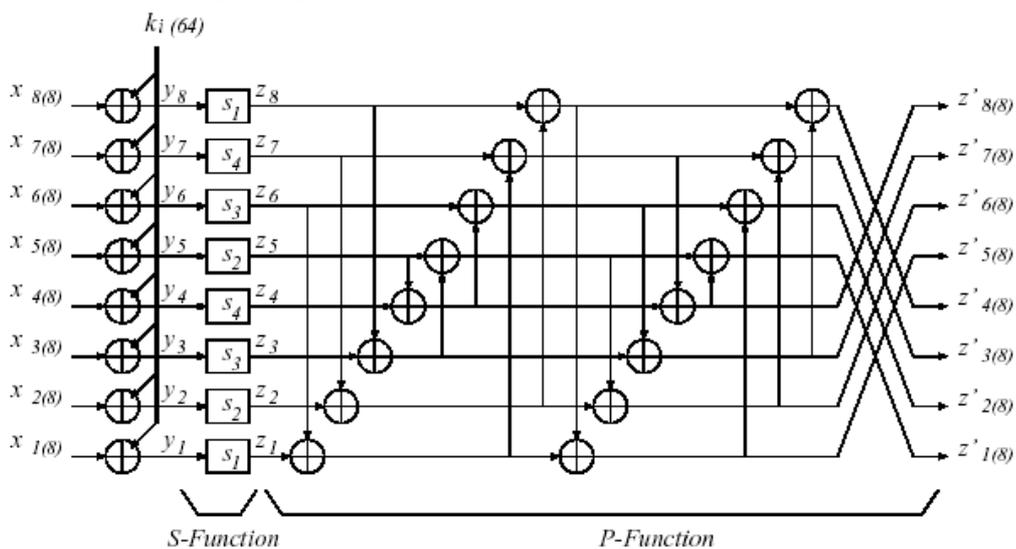

**Fig. 4.2: F-function**

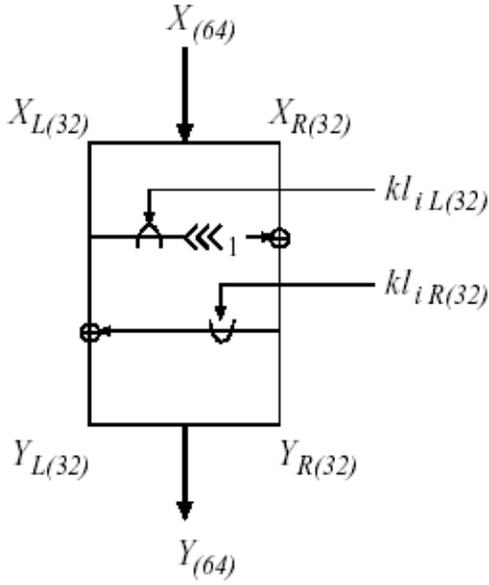 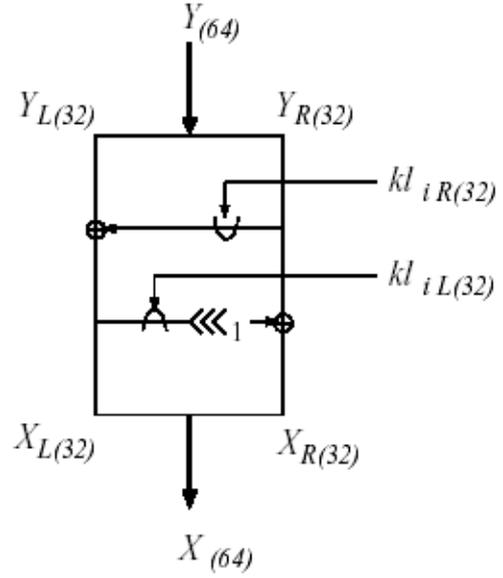

**Fig. 4.3: FL – function**       **Fig. 4.4: FL$^{-1}$ - function**

### 4.1.2 Key Schedule

In the key schedule part of Camellia, we introduce two 128-bit variables $K_{L(128)}$, $K_{R(128)}$ and four 64-bit variables $K_{LL(64)}$, $K_{LR(64)}$, $K_{RL(64)}$ and $K_{RR(64)}$, which are defined in the way that the following relations are satisfactory:

$$K_{(128)} = K_{L(128)} \qquad K_{R(128)} = 0$$

for 128-bit key,

Using these variables, we generate two 128-bit variables $K_{A(128)}$ and $K_{B(128)}$, as shown in Figure 8, where $K_{B(128)}$ is used only if the length of the secret key is 192 or 256 bits. First K = $K_{L(128)}$ is XORed with $K_{R(128)}$ and "encrypted" by two rounds using the constant values $\Sigma_{1(64)}$ and $\Sigma_{2(64)}$ as "keys". The result is XORed with $K_{L(128)}$ and again encrypted by two rounds using the constant values $\Sigma_{3(64)}$ and $\Sigma_{4(64)}$. The resultant value is $K_{A(128)}$. Lastly $K_{A(128)}$ is XORed with $K_{R(128)}$ and encrypted by two rounds using the constant values $\Sigma_{5(64)}$ and $\Sigma_{6(64)}$, the resultant value is $K_{B(128)}$.

The subkeys $kw_{t(64)}$, $k_{u(64)}$, and $kl_{v(64)}$ are generated from (left-half or right-half part of) rotate shifted values of $K_{L(128)}$, $K_{R(128)}$, $K_{A(128)}$, and $K_{B(128)}$.

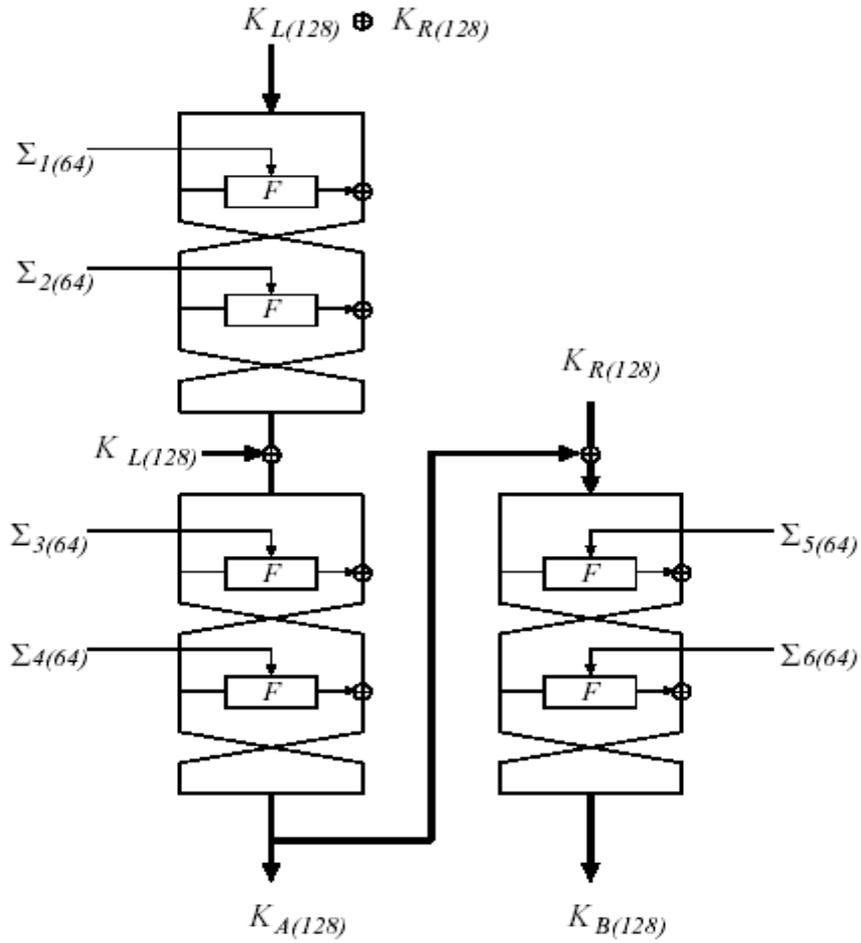

**Fig. 4.5 : Key Schedule**

| $\Sigma_{1(64)}$ | 0xA09E667F3BCC908B |
| $\Sigma_{2(64)}$ | 0xB67AE8584CAA73B2 |
| $\Sigma_{3(64)}$ | 0xC6EF372FE94F82BE |
| $\Sigma_{4(64)}$ | 0x54FF53A5F1D36F1C |
| $\Sigma_{5(64)}$ | 0x10E527FADE682D1D |
| $\Sigma_{6(64)}$ | 0xB05688C2B3E6C1FD |

**Table 4.1: Constants**

|  | subkey | value |
|---|---|---|
| Prewhitening | $kw_{1(64)}$ | $(K_L \lll _0)_{L(64)}$ |
|  | $kw_{2(64)}$ | $(K_L \lll _0)_{R(64)}$ |
| $F$ (Round1) | $k_{1(64)}$ | $(K_A \lll _0)_{L(64)}$ |
| $F$ (Round2) | $k_{2(64)}$ | $(K_A \lll _0)_{R(64)}$ |
| $F$ (Round3) | $k_{3(64)}$ | $(K_L \lll _{15})_{L(64)}$ |
| $F$ (Round4) | $k_{4(64)}$ | $(K_L \lll _{15})_{R(64)}$ |
| $F$ (Round5) | $k_{5(64)}$ | $(K_A \lll _{15})_{L(64)}$ |
| $F$ (Round6) | $k_{6(64)}$ | $(K_A \lll _{15})_{R(64)}$ |
| $FL$ | $kl_{1(64)}$ | $(K_A \lll _{30})_{L(64)}$ |
| $FL^{-1}$ | $kl_{2(64)}$ | $(K_A \lll _{30})_{R(64)}$ |
| $F$ (Round7) | $k_{7(64)}$ | $(K_L \lll _{45})_{L(64)}$ |
| $F$ (Round8) | $k_{8(64)}$ | $(K_L \lll _{45})_{R(64)}$ |
| $F$ (Round9) | $k_{9(64)}$ | $(K_A \lll _{45})_{L(64)}$ |
| $F$ (Round10) | $k_{10(64)}$ | $(K_L \lll _{60})_{R(64)}$ |
| $F$ (Round11) | $k_{11(64)}$ | $(K_A \lll _{60})_{L(64)}$ |
| $F$ (Round12) | $k_{12(64)}$ | $(K_A \lll _{60})_{R(64)}$ |
| $FL$ | $kl_{3(64)}$ | $(K_L \lll _{77})_{L(64)}$ |
| $FL^{-1}$ | $kl_{4(64)}$ | $(K_L \lll _{77})_{R(64)}$ |
| $F$ (Round13) | $k_{13(64)}$ | $(K_L \lll _{94})_{L(64)}$ |
| $F$ (Round14) | $k_{14(64)}$ | $(K_L \lll _{94})_{R(64)}$ |
| $F$ (Round15) | $k_{15(64)}$ | $(K_A \lll _{94})_{L(64)}$ |
| $F$ (Round16) | $k_{16(64)}$ | $(K_A \lll _{94})_{R(64)}$ |
| $F$ (Round17) | $k_{17(64)}$ | $(K_L \lll _{111})_{L(64)}$ |
| $F$ (Round18) | $k_{18(64)}$ | $(K_L \lll _{111})_{R(64)}$ |
| Postwhitening | $kw_{3(64)}$ | $(K_A \lll _{111})_{L(64)}$ |
|  | $kw_{4(64)}$ | $(K_A \lll _{111})_{R(64)}$ |

**Table 4.2: Generation of subkeys**

## 4.2 Analysis of the CAMELLIA main components

### 4.2.1 Substitution boxes

One round of CAMELLIA has got only 8 sboxes and it is possible to implement max. three rounds in one clock cycle, if we want 24 sboxes in EABs. It is possible to prepare implementations:

- 1 clock cycle – 1 round of encryption,
- 1 clock cycle – 2 rounds of encryption,
- 1 clock cycle – 3 rounds of encryption,

when we implement sboxes in EABs.

It turned out that the best solution is the third case.

### 4.2.2 P – function

This kind of operation is similar to operations described in section 2.1.3 (MDS higher level in HIEROCRYPT-3) in implementation (Fig. 4.2).

### 4.2.3 FL – function

FL – function is defined as follow (Fig.4.3):

$$FL : \mathbf{L} \times \mathbf{L} \longrightarrow \mathbf{L}$$
$$(X_{L(32)} || X_{R(32)}, kl_{L(32)} || kl_{R(32)}) \longmapsto Y_{L(32)} || Y_{R(32)},$$

detailed:

$$Y_{R(32)} = ((X_{L(32)} \cap kl_{L(32)}) \lll 1) \oplus X_{R(32)},$$
$$Y_{L(32)} = (Y_{R(32)} \cup kl_{R(32)}) \oplus X_{L(32)}.$$

There are four kind of operations:
- logical AND,
- logical OR,
- shift by 1 bit left,
- EXOR.

All of these operation are hardware oriented and implementation of FL-function is very simple.

### 4.2.4 $FL^{-1}$ - function

Implementation of this operation is as easy as the previous (Fig.4.3).

## 4.3 Implementation of the CAMELLIA and its results

### 4.3.1 Performance of CAMELLIA (Hardware Performance)

Performance of CAMELLIA algorithm is given in [9]. The table is from this paper and it describes architectures and its throughput.

| Architecture | Design library | Throughput |
| --- | --- | --- |
| Loop | Xilinx XC4000XL | 77,34 Mb/s |
|  | Xilinx VirtexE | 199,46 Mb/s |
|  | Xilinx VirtexE | 211,90 Mb/s |
|  | Xilinx VirtexE | **227, 42 Mb/s** |
| Unrolled | Xilinx VirtexE | 401,89 Mb/s |
| Pipeline | Xilinx VirtexE | 6749,99 Mb/s |

Table 4.3: FPGA hardware performance of CAMELLIA.

### 4.3.2 Proposition of implementation of CAMELLIA.

The most satisfactory results of implementation of CAMELLIA algorithm are achieved using loop-unrolled architecture. It means that in one clock cycle we execute 3 rounds of encryption (decryption).

We used the same interface as in HIEROCRYPT-3 projects. It is shown Figure 3.1.
**SETUP phase** – executed in 2 clock cycles:
1st clock cycle:
key schedule – part I. (execution of 2 rounds and the output of these two rounds is EXORed with $K_{L(128)}$),
2nd clock cycle:
key schedule – part II (execution of 2 next rounds).

**WORK phase** – executed in 6 clock cycles:
1st clock cycle:
encryption (decryption) – part I (execution of pre-whitening operation and 3 rounds),

2nd clock cycle:

encryption (decryption) – part II (execution of 3 rounds and FL, FL$^{-1}$ operation),

3rd clock cycle:

encryption (decryption) – part III (execution of 3 rounds),

4th clock cycle:

encryption (decryption) – part IV (execution of 3 rounds and FL, FL$^{-1}$ operation),

5th clock cycle:

encryption (decryption) – part V (execution of 3 rounds),

6th clock cycle:

encryption (decryption) – part VI (execution of 3 rounds and post-whitening operation),

The division of this procedure provides that the round of encryption (decryption) is critical path. Frequency of the clock could be 13,15 MHz and throughput of this project is **240 Mb/s**. Full LOOP – architecture achieved 227 Mb/s, Full UNROLLED – architecture 401 Mb/s. Some papers [14] prove the theorem that the solution called LOOP-UNROLLED – architecture achieves the results of throughput between the best result of LOOP and UNROLLED – architectures. Our results confirm this theorem.

The efficiency of CAMELLIA implementation is 2973 logic elements and 49152 memory bits.

## 5. Conclusions

The implementation of HIEROCRYPT-3 is not simple. The optimal implementation of this algorithm is achieved when all conditions from section 3.5 are taken seriously. This implementation has a very high operation speed **304 Mb/s** and it is almost 6 times faster than the fastest implementation proposed by the authors. This proposition of implementation needs only 9758 logic elements and 48 kb of EAB (embedded array block) - additional memory, it is twice more efficient than that proposed by the authors and it fits to one FPGA circuit.

HIEROCRYPT-3 is a very flexible algorithm. It is possible to connect substitution layer with MDS lower level layer and replace them by one substitution layer with 64 sboxes and few xor-operations. This project needs a lot of logic elements (more than 25000 logic elements), but it is still a practical implementation and its performance is **397 Mb/s**.

It is easy to implement CAMELLIA in hardware. We achieve the best result of throughput when we execute three rounds in one clock cycle (**240 Mb/s**). We call this project LOOP-UNROLLED architecture.

Both ciphers seem to be very suitable for hardware implementation, but, surprisingly, we achieved better results of throughput for HIEROCRYPT-3. However, as to efficiency CAMELLIA is still better.

Our work suggests that possibilities of the algorithm's implementation (HIEROCRYPT-3) should not be evaluated by authors only who very often have not enough knowledge about optimalisation in designing.

At the end of our paper we present comparison of presented implementation by the authors of the primitives: HIEROCRYPT-3 and CAMELLIA and our projects.

**HIEROCRYPT-3:**

|  | efficiency | throughput |
|---|---|---|
| Our proposal: Project with short setup | 8599 LE / 48kb EAB | 115 Mb/s |
| Our proposal: Project with long setup | 9497 LE / 48kb EAB | 190 Mb/s |
| Our proposal: Project with very long setup | 9758 LE / 48kb EAB | 304 Mb/s |
| Our proposal: Extensive solution | 25811 LE | 397 Mb/s |
| TOSHIBA Corp.:high speed | 22700 LE | 52,6 Mb/s |
| TOSHIBA Corp.:small area | 6300 LE | 4,1 Mb/s |

**Table 5.1:Results of implementation of HIEROCRYPT-3**

It is difficult to compare the results of efficiency in CAMELLIA implementation because of the differences in technology between Xilinx and ALTERA circuits.

**CAMELLIA:**

|  | efficiency | throughput |
|---|---|---|
| Our proposal | 2973 LE / 48kb EAB | 240 Mb/s |
| NTT & Mitshubishi EC: Loop | 1296 LE | 77,34 Mb/s |
| NTT & Mitshubishi EC: Loop | 1816 LE | 199,46 Mb/s |
| NTT & Mitshubishi EC: Loop | 1816 LE | 211,90 Mb/s |
| NTT & Mitshubishi EC: Loop | 1780 LE | 227, 42 Mb/s |
| NTT & Mitshubishi EC: Unrolled | 9426 LE | 401,89 Mb/s |

**Table 5.2: Results of implementation of CAMELLIA.**

## Acknowledgements

I would like to thank Piotr BORA for valuable comments and discussions on the implementation of the ciphers: Hierocrypt-3 and Camellia.